\documentclass{article}
\newtheorem{Theorem}{Theorem}
\newtheorem{Lemma}[Theorem]{Lemma}
\newtheorem{Corollary}[Theorem]{Corollary}

\newtheorem{Claim}[Theorem]{Claim}
\begin{document}
\title{Agreement Protocols on an Arbitrary Network in the Presence of a Mobile Adversary}
\author{Chris Dowden\thanks{Author's Note: The original version of this work was wriiten in 2012. For more up-to-date results,
see for example `Tight Bound on Mobile Byzantine Agreement'
by Bonnet, D\'efago, Nguyen and Potop-Butucaru.}}
\date{}
\maketitle
\setlength{\unitlength}{1cm}

\begin{abstract}
We investigate the problem of obtaining agreement protocols in the presence of a mobile adversary,
who can control an ever-changing selection of processors.
We make improvements to previous results for the case when the communications network forms a complete graph,
and also adapt these to the general case when the network is not complete.
\end{abstract}

\section{Introduction} \label{intro}

Suppose various processors in a network wish to reach agreement on a particular decision.
Unfortunately, some unknown subset of these may be under the control of a malicious adversary
who desires to prevent such an agreement being possible.

To this end, the adversary will instruct his `faulty' processors to provide inaccurate information to the non-faulty processors in an attempt to mislead them.
The aim is to construct an `agreement protocol' that will always foil the adversary and enable the non-faulty processors to reach agreement successfully
(perhaps after several rounds of communication).

In traditional agreement problems,
it is usually assumed that the set of faulty processors is `static',
in the sense that it is chosen by the adversary at the start of the process 
and then remains fixed throughout all communication rounds.
In this paper,
we shall instead focus on the case of a `mobile' adversary,
who can continually change his selection of faulty processors.
We assume that the communication links between processors are always perfectly reliable,
and cannot themselves be controlled by the adversary.

This mobile problem was previously investigated in \cite{Reischuk:1985},
for the particular scenario of a complete communications network,
and a successful protocol was given for the case when the total number of processors $n$
and the number of mobile faulty processors $m$ satisfy $n>18m$.
We shall improve on that bound,
as well as generalising our results to the non-complete case.

In the remainder of this section,
we describe the basic set-up,
and summarise the static case.
In Section~\ref{mobile}, 
we then introduce the mobile version,
and provide details of relevant results;
in Section~\ref{latest},
we present a new algorithm for the specific scenario of mobile adversaries on a complete network,
and reduce the $n>18m$ requirement to just $n>6m$;
in Section~\ref{intro2},
we investigate how to adapt this procedure to arbitrary networks;
and in Section~\ref{apps},
we consequently derive conditions in terms of the vertex-connectivity and minimum degree of such networks.

\subsection{Formulation of the problem} \label{formulation}

Throughout this paper,
we shall consider the situation where there are $n$ processors (i.e.~vertices)
connected by a network of links (i.e.~edges) enabling pairwise communication,
but where some of the processors are faulty.
We assume that the communication links themselves are always perfectly reliable.

Crucially, the exact identity of the faulty processors is not known,
but the overall aim is for the non-faulty processors to nevertheless reach some sort of `agreement' about a particular decision,
by adhering to a pre-determined protocol.

In this paper,
we shall concentrate on the `Byzantine Agreement' (BA) formulation of the agreement problem.
Here,
a known source processor $s$ (which may or may not be one of the faulty ones)
is supposed to transmit a value $v_{s}$ over all its communication links.
If $s$ is non-faulty then it will indeed follow these instructions,
but if $s$ is faulty then it could send anything ---
in particular, it might send different values to different processors.

All the processors (including $s$) are then permitted to communicate with each other over the network for as long as they desire ---
note that the faulty processors will use this as an opportunity to spread inaccurate information.
The aim is for the non-faulty processors to eventually reach agreement about the value of $v_{s}$ ---
we shall define the exact form that this agreement must take later.

We shall assume throughout that all processors,
faulty and non-faulty,
are always completely synchronised.
Hence, each `communication round' will consist of
(i)~all processors simultaneously sending out various messages to adjacent processors,
and then (ii)~each processor receiving the messages sent to it during that round.

Note that we shall also always assume a `non-authenticated' environment,
where the faulty processors can relay information incorrectly,
rather than the `authenticated' version,
where the faulty processors can merely refuse to pass on information.

\subsection{The static adversary} \label{static}

In this subsection,
we shall describe the standard `static' case,
where the set of faulty processors is chosen by the adversary at the start of the process and then remains fixed.

Recall that we are concentrating on the BA formulation of the agreement problem,
where the source processor $s$ starts with a value $v_{s}$.

In the static case,
a successful protocol should eventually terminate with the non-faulty processors having reached an agreement about the value of $v_{s}$ that satisfies the following two conditions: \\
\\
(BA1) every non-faulty processor must agree on a common value $v_{s}^{\prime}$; \\
(BA2) if the source processor is non-faulty,
then $v_{s}^{\prime}$ must be the correct value $v_{s}$. \\

The basic case,
first studied in \cite{Pease:1980},
is when the communications network forms a complete graph
(meaning that every pair of processors can communicate directly),
and when the adversary can control any $m$ of the $n$ processors, for some fixed $m \leq n-2$.
We assume that the value of $m$ is common knowledge,
but that the specific identity of the controlled processors is known only to the adversary.

In this setting,
it is known that it is then possible to construct a successful agreement protocol for the BA problem
if and only if $n>3m$
 (see \cite{Pease:1980} for a clever constructive proof by induction on $m$,
which essentially works by demonstrating that if there is much contradictory information
then $s$ must be faulty,
in which case the induction hypothesis can be used on the remaining processors).

In the non-complete case,
some processors will not be able to communicate directly with one another,
and will instead have to go through third parties.
In particular,
this means that any message sent by processor $p$ to processor $q$
will take at least $d(p,q)$ communication rounds to reach its destination,
where $d(p,q)$ denotes the distance between $p$ and $q$,
and may be tampered with along the way.

However,
it is simple to observe
(as in \cite{Dolev:1993}, for example)
that every pair of processors can still communicate effectively as long as the network is $2m+1$ vertex-connected
(simply transmit any desired message along $2m+1$ pre-agreed vertex-disjoint paths ---
the majority of these messages will arrive unaltered).
Consequently,
it is possible to construct a successful agreement protocol if and only if
$n>3m$
and the network is $2m+1$ vertex-connected.

In addition to Byzantine agreement,
there are also two other common formulations of the agreement problem ---
`consensus' and `interactive consistency'.
The definitions differ in terms of the initial set-up and the form of the final agreement,
but it can be shown that all three formulations are actually equivalent,
in the sense that a protocol to solve any one of the problems can be modified into protocols to also solve the other two.

\section{Introduction to the mobile adversary} \label{mobile}

In the previous subsection,
we outlined some results for the case of a static adversary.
In this section,
we shall now start to instead explore the concept of a mobile adversary,
which is to be the focus of this paper.
In Subsection~\ref{mobileintro},
we provide a brief introduction;
in Subsection~\ref{comm},
we state the assumed communication procedure;
in Subsection~\ref{strong},
we shall look at one way to formulate the problem,
which results in no successful protocols being possible;
and in Subsection~\ref{weak},
we discuss a more promising approach.
In Subsection~\ref{other},
we also outline a few miscellaneous results.

Existing results on mobile adversaries seem to concentrate on the case when the communications network is complete.
In Section~\ref{latest},
we shall present a new algorithm for this case,
before moving on to general networks in Sections~\ref{intro2} and~\ref{apps}.

Let us re-iterate here that we are concerned with an adversary who can control processors,
but not communication links.
There is already a substantial body of work
(see, for example, \cite{Santoro:2007}, \cite{Schmid:2009} and the references therein)
dealing with the problem of mobile link faults,
but this is not of direct relevance to us.

\subsection{Introduction} \label{mobileintro}

In contrast to the static case considered in the previous section,
we shall now allow the adversary to change his $m$ selected processors at the start of each new communication round.
In particular, there is to be no bound on the total number of processors that may be affected at some stage during the process.

There are various different ways that the basic Byzantine agreement problem can then be re-stated.
For example, the precise formulation will depend on issues such as who exactly is required to reach agreement
(should it be all processors who are non-faulty at that particular moment or just all processors who have never been faulty?
--- we shall discuss this in Subsections~\ref{strong} and~\ref{weak})
and to what extent (if any) the adversary is allowed to tamper with the memory of a faulty processor.

In terms of the latter issue,
one option that appears natural,
and which we shall henceforth choose,
is to allow the adversary \emph{unlimited} ability to rewrite any data stored by the processors currently under its control
(note that this means that even after a processor is deselected, 
it will still retain potentially inaccurate information stored by the adversary),
including any data recorded prior to becoming faulty.

Of course,
we shall also still assume that when under the control of the adversary,
a processor may completely disregard the pre-determined protocol ---
in particular,
by sending inaccurate and inconsistent messages to the other processors.
However,
we shall suppose that the protocol itself is stored safely and cannot be tampered with.
Consequently,
if deselected by the adversary,
a processor will then return to following the correct protocol again
(albeit possibly with inaccurate data stored by the adversary).

Throughout the rest of this paper,
we shall use the term `faulty' to refer to the $m$ processors which are currently under the control of the adversary,
and `non-faulty' to refer to the other $n-m$ processors,
including any which have been controlled previously.

\subsection{The communication procedure} \label{comm}

We shall assume that each communication round proceeds as follows: \\
\\
\textbf{Step $\mathbf{1}$}: The adversary chooses which $m$ processors to control during this round
(he will not be able to change his selection again until the start of the next round);
each processor starts the round with the stored information that it has collated during previous rounds
(if the processor was controlled at some point by the adversary,
then some of this information may be incorrect); \\
\textbf{Step $\mathbf{2}$}: Each non-faulty processor sends out messages to all adjacent processors,
based on the state of its stored information and the instructions of the protocol;
the processors that are currently under the control of the adversary may send out anything in their messages; \\
\textbf{Step $\mathbf{3}$}: Every processor receives the messages that were sent to it during Step~$2$; \\
\textbf{Step $\mathbf{4}$}: Each non-faulty processor updates its stored information according to the instructions of the protocol,
to take into account the messages sent and received in Steps $2$ and $3$;
the adversary may store incorrect information on the processors currently under its control,
\emph{and may even maliciously rewrite information that was already stored}. \\

A successful protocol should eventually terminate
(after some pre-determined fixed number of rounds,
independent of the actions of the adversary)
with the relevant processors having reached a suitable agreement
(see Subsections~\ref{strong} and~\ref{weak} for the precise requirements).

\subsection{Impossibly strong agreement requirements} \label{strong}

Recall that we are concentrating on the BA formulation of the agreement problem,
where the source processor $s$ starts with a value $v_{s}$.
As already noted,
there are different possibilities for how to rephrase BA1 and BA2 in the case of a mobile adversary.
In this subsection,
we shall look at what happens if we require that the agreement must satisfy the following two conditions: \\
\\
(MBA1) every processor which is non-faulty \emph{in the final round} must agree on a common value $v_{s}^{\prime}$; \\
(MBA2) if the source processor has \emph{never} been faulty,
then $v_{s}^{\prime}$ must be the correct source value $v_{s}$. 
\\

Unfortunately,
(a version of) this problem was examined in \cite{Garay:1994},
and it was observed that it is actually impossible to construct a successful protocol that always satisfies these two conditions,
even when the communications network is complete and $m=1$!:

\begin{Theorem} [rephrased from \cite{Garay:1994}, Proposition 1] \label{imposs}
Let $m \geq 1$ denote the number of mobile faulty processors in each communication round,
and let $n>m+1$ denote the total number of processors.
Then it is impossible to construct a protocol for non-authenticated mobile Byzantine agreement that always satisfies conditions MBA1 and MBA2.
\end{Theorem}

The proof is based on work in \cite{Dolev:1983} and \cite{Fischer:1982}, 
and involves making a series of minor modifications to the behaviour of faulty processors in intermediate rounds,
working back earlier and earlier without changing the value of the final agreement
until eventually $v_{s}$ itself is altered
and condition MBA2 is contradicted.
A crucial ingredient is the fact that MBA1 implies that all processors non-faulty in the final round need to agree 
even if each was faulty in an intermediate round.

\subsection{Weaker agreement requirements} \label{weak}

Due to the impossibility result of Theorem~\ref{imposs},
it is more interesting to look at what happens if condition MBA1 is weakened to just require agreement from all processors that are never faulty
(with condition MBA2 remaining the same),
i.e.: \\
\\
(MBA$1^{\prime}$) every processor which has \emph{never} been faulty must agree on a common value $v_{s}^{\prime}$; \\
(MBA$2$) if the source processor has never been faulty,
then $v_{s}^{\prime}$ must be the correct source value $v_{s}$. \\

With these weaker agreement requirements,
a successful protocol for the case when the network is complete and $n > 18m$
then follows as a corollary to the following detailed
result of Reischuk (\cite{Reischuk:1985}),
where rounds are grouped in `intervals' of three:

\begin{Theorem} \label{Reischuk} \textbf{(\cite{Reischuk:1985}, Theorem $\mathbf{1}$)}
Let the network be complete,
let $n$ denote the total number of processors,
let $p_{1}, p_{2}, \ldots, p_{n}$ denote the processors themselves,
with $p_{1}$ being the source processor $s$,
and let $t< \frac{n}{6}$.
Suppose that (i)~in each $3$-round interval,
at most $t$ processors are faulty in at least one of these rounds,
and (ii)~there exists $R$ such that processor $p_{R}$ is non-faulty in rounds $2R$ and $2R+1$.

Then there exists a protocol for non-authenticated mobile Byzantine agreement containing a special round $r^{*} \leq 2R+3$ for which 
(a) all processors which are non-faulty in both rounds $r^{*}-1$ and $r^{*}$ will have set the same value of $v_{s}^{\prime}$ by round $r^{*}$
and will not change it as long as they stay non-faulty,
and (b) if processor $p_{i}$ is non-faulty in both rounds $r^{*}-1$ and $r^{*}$ and the source processor is non-faulty in round $1$,
then $p_{i}$'s value of $v_{s}^{\prime}$ at round $r^{*}$ will be the correct source value $v_{s}$.
\end{Theorem}

\begin{Corollary} \label{Corollary}
Let the network be complete,
let $m$ denote the number of mobile faulty processors in each communication round,
and let $n>18m$ denote the total number of processors.
Then there exists a protocol for non-authenticated mobile Byzantine agreement that always satisfies conditions MBA$1^{\prime}$ and MBA$2$.
\end{Corollary}
\textbf{Proof}
Observe that requirement MBA$1^{\prime}$ is vacuous unless there exists a processor $p_{R}$ that is always non-faulty,
and hence we may apply Theorem~\ref{Reischuk} with $t=3m$.
\phantom{qwerty}
\setlength{\unitlength}{.25cm}
\begin{picture}(1,1)
\put(0,0){\line(1,0){1}}
\put(0,0){\line(0,1){1}}
\put(1,1){\line(-1,0){1}}
\put(1,1){\line(0,-1){1}}
\end{picture} \\

As an aside,
note that Theorem~\ref{Reischuk} actually provides a successful protocol even if we strengthen condition MBA2 to
require that $v_{s}^{\prime} = v_{s}$ whenever $s$ is non-faulty in round $1$,
rather than just whenever $s$ is always non-faulty.

One of the main difficulties with agreement problems is the fact that one processor may believe
that the support for a particular value of $v_{s}^{\prime}$ is just above a given threshold,
while another processor receiving slightly different information may instead think
that the amount of support is just below the threshold.
Reischuk's successful protocol gets around this issue by instead introducing two separate threshold levels,
essentially classifying the degree of support into `high', `medium' or `low'.
The crucial ingredient in the protocol
is the role then played by a special processor
(whose identity changes every two rounds),
which uses its own view of the degree of support to encourage all `medium' classifications to crystallise
into either `high' or `low',
thus enabling a decision to be reached.

In Section~\ref{latest},
we shall present a refined version of Reischuk's protocol,
and improve on the bounds given by Theorem~\ref{Reischuk}/Corollary~\ref{Corollary}.
In Sections~\ref{intro2} and~\ref{apps},
we shall then adapt this protocol to the case of non-complete networks.

\subsection{Alternative formulations} \label{other}

Various other slightly different formulations of the mobile adversary agreement problem have also been discussed,
again concentrating on the case when the network is complete.

In \cite{Banu:2012} and \cite{Garay:1994},
it is assumed that any processor changing from faulty to non-faulty will immediately know that it was previously faulty.
In this setting, a successful protocol is given in \cite{Banu:2012} for $n > 4m$.

In \cite{Biely:2011},
a version of the problem is investigated 
under the additional assumption that no new processors can fail 
before recovered processors have successfully learned the current state of computation.
With this modification,
it is observed that successful protocols are possible for $n>3m$.

In \cite{Buhrman:1995},
a slightly different formulation of the problem is proposed,
where the adversary chooses his selection of faulty processors between Step $2$ and Step $3$, instead of during Step $1$.
Crucially, this means that any processor released from the control of the adversary is able to receive and \emph{then} send out messages 
before it can be put under the control of the adversary again.
Here,
a successful protocol is given for $n > 3m$.

\section{Complete networks} \label{latest}

In this section,
we now present new results for the formulation of the problem outlined in Subsection~\ref{weak},
for the specific case when the network is complete.
In Subsection~\ref{protocol}, we refine Reischuk's protocol;
in Subsection~\ref{suff}, we use this to improve the $n>18m$ result of Corollary~\ref{Corollary} to just $n>6m$;
and in Subsection~\ref{nec}, we observe that $n>5m$ is certainly necessary.
In the remainder of the paper,
we shall then look at generalising our results to non-complete networks.

\subsection{The protocol} \label{protocol}

The protocol is based on Reischuk's (\cite{Reischuk:1985}),
with each processor $p_{i}$ using two sets $A_{i}$ and $B_{i}$ 
to contain the set of possible values of $v_{s}$ that he believes to have `high' or `medium' support from the other processors. 
In terms of improving the results of Theorem~\ref{Reischuk} and Corollary~\ref{Corollary},
the crucial modification that we make is for $p_{i}$ to continue transmitting information about $A_{i}$ and $B_{i}$ to the other processors even after setting his value of $v_{s}^{\prime}$,
thus enabling the other processors to gain extra information.

Additionally,
in order to cut down on the amount of data that is transmitted in each communication round
(and hence reduce time and cost),
we introduce two new values $\bot_{0}$ and $\bot_{2}$
for the cases when the sets $A_{i}$ and $B_{i}$ have size zero or size at least two,
and we consequently modify the protocol so that it is just two values $a_{i}$ and $b_{i}$ that are transmitted by each processor,
rather than the entire sets $A_{i}$ and $B_{i}$.

To simplify some of the expressions in this new version of the protocol,
and the subsequent proof,
we also alter the definition of $B_{i}$ so that it is always a superset of $A_{i}$,
rather than being disjoint,
and we allow both $A_{i}$ and $B_{i}$ to contain the value $\bot_{2}$
(but not the value $\bot_{0}$).

Without loss of generality,
we denote the $n$ processors by $p_{1}, p_{2}, \ldots, p_{n}$,
where the source processor $s$ is $p_{1}$. 

As mentioned,
the protocol presented here is for the case when the network is complete.
The non-complete case will be discussed in Sections~\ref{intro2} and~\ref{apps}. \\
\\
\textbf{ROUND} $\mathbf{1}$ \\
\textbf{Transmission:}
The source processor $s=p_{1}$ sends $v_{s}$ to all processors (including himself). \\
\textbf{Setting $\mathbf{a_{l}}$ \textbf{and} $\mathbf{b_{l}}$:}
Each processor $p_{l}$ then sets both $a_{l}$ and $b_{l}$ to be equal to the purported value of $v_{s}$ that he has just received from the source processor
(the values of $a_{l}$ and $b_{l}$ will be modified at the end of each subsequent round). \\
\\
\textbf{ROUND $\mathbf{r}$, FOR $\mathbf{r \in \{2,3, \ldots, 2n \}}$} \\
\textbf{Transmission:}
Each processor $p_{i}$ sends the current values of $a_{i}$ and $b_{i}$ to all processors (including himself). \\
\textbf{Setting/resetting $\mathbf{v_{s}^{\prime}}$:}
Let $a_{ij}$ and $b_{ij}$ denote the values received by processor $p_{j}$ from processor $p_{i}$. 
In the event that all but at most $2m$ of the values $a_{1l}, a_{2l}, \ldots, a_{nl}$ are equal to a common value $a$,
processor $p_{l}$ sets/resets his value of $v_{s}^{\prime}$ to be $a$.
Otherwise, he leaves his value of $v_{s}^{\prime}$ unset/unchanged. \\
\textbf{Setting/resetting $\mathbf{A_{l}}$ \textbf{and} $\mathbf{B_{l}}$:}
Let $f(r)= \lfloor \frac{r}{2} \rfloor + 1$. 
For $l \neq f(r)$,
processor $p_{l}$ sets/resets $A_{l}$ so that $x \in A_{l}$ if and only if $x \neq \bot_{0}$ and either \\
(a) $a_{f(r)l}=x$ and $b_{jl} \in \{x, \bot_{2} \}$ for strictly more than $4m$ values of $j$ in total \\
or (b) $a_{jl}=x$ for strictly more than $4m$ values of $j$ in total; \\
and he sets/resets $B_{l}$ so that $x \in B_{l}$ if and only if $x \neq \bot_{0}$ and either \\
(a) $a_{f(r)l}=x$ and $b_{jl} \in \{x, \bot_{2} \}$ for strictly more than $2m$ values of $j$ in total \\
or (b) $a_{jl}=x$ for strictly more than $2m$ values of $j$ in total. 

The `special' processor $p_{f(r)}$ sets/resets $A_{f(r)}$ so that $x \in A_{f(r)}$ if and only if $x \neq \bot_{0}$ and either \\
(a) $a_{f(r)f(r)}=x$ and $b_{jf(r)} \in \{x, \bot_{2} \}$ for strictly more than $3m$ values of $j$ in total \\
or (b) $a_{jf(r)}=x$ for strictly more than $3m$ values of $j$ in total; \\
and he sets/resets $B_{f(r)}$ to be equal to this new $A_{f(r)}$. \\
\textbf{Resetting $\mathbf{a_{l}}$ \textbf{and} $\mathbf{b_{l}}$:}
Each processor $p_{l}$ (including the case $l=f(r)$) resets $a_{l}$ and $b_{l}$ so that
\begin{displaymath}
a_{l} =
\left\{ \begin{array}{ll}
x & \textrm{if $A_{l} = \{ x \}$ (including the case when $x=\bot_{2}$)} \\
\bot_{0} & \textrm{if $|A_{l}|=0$} \\
\bot_{2} & \textrm{if $|A_{l}| \geq 2$} \\
\end{array} \right. 
\end{displaymath} 
\begin{displaymath}
\textrm{ and } b_{l} =
\left\{ \begin{array}{ll}
x & \textrm{if $B_{l} = \{ x \}$ (including the case when $x=\bot_{2}$)} \\
\bot_{0} & \textrm{if $|B_{l}|=0$} \\
\bot_{2} & \textrm{if $|B_{l}| \geq 2$.} \\
\end{array} \right. 
\end{displaymath} \\
\\
\textbf{TERMINATION} \\
The protocol terminates after $2n$ rounds.

\subsection{Sufficiency of $\mathbf{n>6m}$} \label{suff}

We shall now present properties of the protocol defined in Subsection~\ref{protocol},
improving on Theroem~\ref{Reischuk} and Corollary~\ref{Corollary}.

\begin{Theorem} \label{6m}
Let the network be complete,
let $m$ denote the number of mobile faulty processors in each communication round,
let $n>6m$ denote the total number of processors,
and let $p_{1}, p_{2}, \ldots, p_{n}$ denote the processors themselves,
with $p_{1}$ being the source processor $s$.
Suppose that there exists $R$ such that processor $p_{R}$ is non-faulty in both rounds $2R-2$ and $2R-1$
(the round $2R-2$ condition can be ignored if $R=1$).

Then the protocol for non-authenticated mobile Byzantine agreement defined in Subsection~\ref{protocol} operates such that for every round $r \geq 2R$
(a) all processors which are non-faulty in round $r$ will have a common value for $v_{s}^{\prime}$ at the end of that round
and (b) if the source processor is non-faulty in round $1$ (i.e.~if $R=1$),
then this value of $v_{s}^{\prime}$ will always be the correct source value $v_{s}$.
\end{Theorem}
\textbf{Proof}
First,
consider the case when the source processor $s=p_{1}$ is non-faulty in round $1$ (i.e.~$R=1$).
Recall that every processor $p_{i}$ which is non-faulty in round $1$ will hence set $a_{i} = v_{s}$ and $b_{i} = v_{s}$.
It can then be observed inductively that in each subsequent round $r \geq 2$,
each processor $p_{j}$ which is non-faulty in both rounds $r-1$ and $r$ will send $a_{j} = v_{s}$ and $b_{j} = v_{s}$,
and (since there are at least $n-2m$ of those)
each processor $p_{k}$ which is non-faulty in round $r$ will consequently set 
$v_{s}^{\prime}$ to be $v_{s}$,
set both $A_{k}$ and $B_{k}$ to be $\{ v_{s} \}$,
and set both $a_{k}$ and $b_{k}$ to be $v_{s}$. 

Now consider the case when the source processor $s=p_{1}$ is faulty in round $1$,
but when there exists a processor $p_{R}$ (for some $R \geq 2$) which is non-faulty in both rounds $2R-2$ and $2R-1$.

\begin{Claim} \label{claim}
Suppose that there exists a processor $p_{R}$,
for some $R \geq 2$,
which is non-faulty in both rounds $2R-2$ and $2R-1$.
Then there exists $a^{*}$ such that,
at the end of round $2R-1$,
every processor $p_{i}$ which is non-faulty in round $2R-1$ will set $a_{i}=a^{*}$ and $b_{i}=a^{*}$.
\end{Claim}

It follows from the claim that all but at most $2m$ processors will send $(a^{*}, a^{*})$ in round $2R$,
and so all non-faulty processors in round $2R$ will certainly set $v_{s}^{\prime} = a^{*}$.
It can then again be observed inductively that in each subsequent round $r \geq 2R$,
each processor $p_{j}$ which is non-faulty in both rounds $r-1$ and $r$ will send $a_{j} = a^{*}$ and $b_{j} = a^{*}$,
and (since there are at least $n-2m$ of those)
each processor $p_{k}$ which is non-faulty in round $r$ will consequently 
set $v_{s}^{\prime}$ to be $a^{*}$,
set $A_{k}$ and $B_{k}$ so that
\begin{displaymath}
\left. \begin{array}{cccccl}
A_{k} & = & B_{k}  & = & \{ a^{*}\} & \textrm{ if } a^{*} \notin \{ \bot_{0}, \bot_{2}\} \\
A_{k} & = & B_{k}  & = &  \emptyset & \textrm{ if } a^{*} = \bot_{0} \\
B_{k} & \supset & A_{k}  & \supset  & \{ \bot_{2}\} & \textrm{ if } a^{*} = \bot_{2}, \\
\end{array} \right.
\end{displaymath}
and set both $a_{k}$ and $b_{k}$ to be $a^{*}$. \\
\\
\textbf{Proof of Claim}
The proof will follow from a careful consideration of the workings of the protocol in rounds $2R-2$ and $2R-1$,
utilising the fact that $f(2R-2)=f(2R-1)=R$.
To avoid confusion about the changing state of the various sets and values in these different rounds,
we will now use $A_{l}^{r}$ to denote the state of $A_{l}$ at the end of round $r$,
and we define $B_{l}^{r}$, $a_{l}^{r}$ and $b_{l}^{r}$ analogously.

First,
observe that only at most $m$ of the messages received by any two processors in any given round can differ,
and recall that $f(2R-2)=f(2R-1)=R$.
It then follows from the definitions in the protocol that,
for each processor $p_{k}$ which is non-faulty in round $2R-2$,
we have
$A_{k}^{2R-2} \subset A_{R}^{2R-2} = B_{R}^{2R-2} \subset B_{k}^{2R-2}$
(note that these sets are well-defined,
since $2R-2 \geq 2$).

If $|A_{R}^{2R-2}|=0$,
we thus have $a_{R}^{2R-2} = \bot_{0}$ and $a_{k}^{2R-2} = \bot_{0}$.
Hence,
in round $2R-1$,
processor $p_{R}$ will send $a_{R} = \bot_{0}$,
and each processor $p_{j}$ which is non-faulty in both rounds $2R-2$ and $2R-1$
(there are at least $n-2m$ of these in total, including $p_{R}$)
will send $a_{j} = \bot_{0}$.
Thus,
each processor $p_{i}$ which is non-faulty in round $2R-1$ will consequently set both $A_{i}^{2R-1}$ and $B_{i}^{2R-1}$ to be $\emptyset$,
and hence set both $a_{i}^{2R-1}$ and $b_{i}^{2R-1}$ to be $\bot_{0}$.

If $A_{R}^{2R-2} = \{ x \}$ (here we include the possibility that $x= \bot_{2}$),
then $a_{R}^{2R-2} = x$, $a_{k}^{2R-2} \in \{ \bot_{0}, x \}$ and $b_{k}^{2R-2} \in \{ x, \bot_{2} \}$.
Hence,
in round $2R-1$,
we obtain $A_{i}^{2R-1} = B_{i}^{2R-1} = \{ x \}$,
and consequently $a_{i}^{2R-1} = b_{i}^{2R-1} = x$.

Finally,
if $|A_{R}^{2R-2}| \geq 2$,
then $a_{R}^{2R-2} = b_{R}^{2R-2} = b_{k}^{2R-2} = \bot_{2}$.
Hence,
in round $2R-1$,
we obtain
$\bot_{2} \in A_{i}^{2R-1}$ and $\bot_{2} \in B_{i}^{2R-1}$,
and consequently $a_{i}^{2R-1} = b_{i}^{2R-1} = \bot_{2}$. \\

The proof of the claim completes the proof of the theorem.
\phantom{qwerty}
\setlength{\unitlength}{.25cm}
\begin{picture}(1,1)
\put(0,0){\line(1,0){1}}
\put(0,0){\line(0,1){1}}
\put(1,1){\line(-1,0){1}}
\put(1,1){\line(0,-1){1}}
\end{picture} \\

\begin{Corollary}
Let the network be complete,
let $m$ denote the number of mobile faulty processors in each communication round,
and let $n>6m$ denote the total number of processors.
Then the protocol defined in Subsection~\ref{protocol} always satisfies conditions MBA$1^{\prime}$ and MBA$2$ for non-authenticated mobile Byzantine agreement.
\end{Corollary}
\textbf{Proof}
Recall that requirement MBA$1^{\prime}$ is vacuous unless there exists a processor $p_{R}$ that is always non-faulty.
\phantom{qwerty}
\setlength{\unitlength}{.25cm}
\begin{picture}(1,1)
\put(0,0){\line(1,0){1}}
\put(0,0){\line(0,1){1}}
\put(1,1){\line(-1,0){1}}
\put(1,1){\line(0,-1){1}}
\end{picture} \\

Note that in certain circumstances,
e.g.~if there exists a processor that is always non-faulty,
Theorem~\ref{6m} implies that the protocol works even if we require that all processors non-faulty in the final round need to agree,
as in MBA$1$
(note that this does not contradict Theorem~\ref{imposs},
since such circumstances may not arise).

\subsection{Necessity of $\mathbf{n>5m}$} \label{nec}

We have just seen that conditions MBA$1^{\prime}$ and MBA$2$ can be satisfied if $n>6m$.
Conversely,
we shall finish this section by now showing that it is certainly necessary to have at least $n>5m$:

\begin{Theorem}
Let $m \geq 1$ denote the number of mobile faulty processors in each communication round,
and let $n>m+1$ denote the total number of processors.
Then if $n \leq 5m$,
it is impossible to construct a protocol for non-authenticated mobile Byzantine agreement that always satisfies conditions MBA$1^{\prime}$ and MBA$2$.
\end{Theorem}
\textbf{Proof}
Let us divide the $n$ processors into five sets each having size at most $m$,
and let us denote these sets by  $S$, $A$, $B$, $C$ and $D$,
with the source processor $s \in S$
(we may assume that $A \cup B$ and $C \cup D$ are both non-empty).

Firstly, imagine that the adversary selects the set $S$ in every round,
and that the processors in $S$ always tell those in $A \cup B$ that $v_{s}=0$, 
but always tell those in $C \cup D$ that $v_{s}=1$.
To satisfy condition MBA$1^{\prime}$, 
it will consequently be necessary for processors in $A \cup B$ to agree with processors in $C \cup D$ on a common value,
but it won't matter what that value is.

Secondly, now consider the separate case when $v_{s}=1$ 
and when the adversary repeatedly selects the sets $A$ and $B$ alternately
($A$ in the first communication round,
$B$ in the second round,
$A$ in the third round, etc.).
Observe that, by storing false information,
it is possible for the adversary to guarantee that the messages from processors in $A \cup B$ to processors in $C \cup D$ will always be so as to suggest that $v_{s}=0$
(this is possible because, in each round, 
one of $A$ and $B$ will be under the control of the adversary, 
while the processors in the other will be accidentally relaying false information stored by the adversary in the previous round).
To satisfy conditions MBA$1^{\prime}$ and MBA$2$,
it will be necessary for processors in $C \cup D$ to agree on a common value,
and for that to be the correct source value $v_{s}=1$.

Crucially, note that it will be impossible for the processors in $C \cup D$ to be able to distinguish between the two cases.
Hence, these processors will be forced to decide on the value $1$ in the first scenario as well as in the second.
By using a symmetrical example to the second case,
it can similarly be shown that the processors in $A \cup B$ will be forced to decide on the value $0$.
Hence, agreement will not be reached. 
\phantom{qwerty}
\setlength{\unitlength}{.25cm}
\begin{picture}(1,1)
\put(0,0){\line(1,0){1}}
\put(0,0){\line(0,1){1}}
\put(1,1){\line(-1,0){1}}
\put(1,1){\line(0,-1){1}}
\end{picture} \\

We are left with just a small gap remaining between the necessary and sufficient conditions.

\section{Arbitrary networks --- the key theorem} \label{intro2}

We shall now consider the case when the network of processors is an \emph{arbitrary} graph
(i.e.~not necessarily complete),
starting in Subsection~\ref{impossible} with an observation
concerning the impossibility of successful protocols when the vertex-connectivity is small.

Recall (from Subsection~\ref{static}) 
that the non-complete case for static adversaries can easily be reduced to the complete case,
provided that the network has sufficient connectivity to guarantee that reliable communications are still possible.
In Subsection~\ref{key},
we present a key analogous result for mobile adversaries.

In Section~\ref{apps},
we shall then use this approach to deduce detailed results phrased in terms of the minimum degree and vertex-connectivity of the network.

\subsection{An impossibility result} \label{impossible}

\begin{Theorem} \label{impossible4m}
Let $m$ denote the number of mobile faulty processors in each communication round,
and suppose the network contains a cut-set of size at most $4m$
that doesn't contain the source processor $s$.
Then
it is impossible to construct a protocol for non-authenticated mobile Byzantine agreement that always satisfies conditions MBA$1^{\prime}$ and MBA$2$.
\end{Theorem}
\textbf{Proof} 
Let $G$ denote the network 
and let $X$ denote the given cut-set.
Let us divide $X$ into four disjoint sets $A$, $B$, $C$ and $D$ of size at most $m$ each,
and let $p$ be an arbitrary processor that is in a different component to the source processor $s$ in the graph $G \setminus X$.
Thus, all messages from $s$ to $p$ must pass through processors in either $A \cup B$ or $C \cup D$.

First, let us consider the scenario where the adversary selects the set $A$ in every odd-numbered round 
and then the set $B$ in every even-numbered round.
Hence, in each round,
one of $A$ and $B$ will be actively under the control of the adversary,
while the other will be relaying information stored by the adversary in the previous round.
Thus, it will be possible that $A \cup B$ will always act as if $v_{s}=1$ and $C \cup D$ are faulty.

Similarly, if $v_{s}=1$ and the sets $C$ and $D$ are selected alternately,
it will be possible that $C \cup D$ will always act as if $v_{s}=0$ and $A \cup B$ are faulty.

In order to satisfy condition MBA$2$
(since both $s$ and $p$ are always non-faulty),
it will be necessary for $p$ to somehow determine the correct source value $v_{s}$ in both cases.
However, since the two scenarios will appear identical to $p$,
this will clearly be impossible. 
\phantom{qwerty}
\setlength{\unitlength}{.25cm}
\begin{picture}(1,1)
\put(0,0){\line(1,0){1}}
\put(0,0){\line(0,1){1}}
\put(1,1){\line(-1,0){1}}
\put(1,1){\line(0,-1){1}}
\end{picture}

\subsection{The key theorem} \label{key}

In this subsection,
we shall now present a detailed result (Theorem~\ref{commtoprotocol})
which essentially reduces the remaining problem to one of constructing reliable communication throughout the network.
We also provide (in Corollary~\ref{maincor})
a simplified version of this same result.

In order that we may use the framework of the protocol given in Subsection~\ref{static},
we shall find it helpful if we parameterise communication systems in terms of $T$,
the amount of time taken for a message to reach its destination reliably,
and $K$,
the amount of time that the sender and recipient need to remain non-faulty
(e.g.~for the complete network case,
$T=K=1$).

We start with the full theorem:

\begin{Theorem} \label{commtoprotocol}
Let $m$ denote the number of mobile faulty processors in each communication round,
let $n$ denote the total number of processors,
and let $p_{1}, p_{2}, \ldots, p_{n}$ denote the processors themselves,
with $p_{1}$ being the source processor $s$.

Suppose that there exist $T$ and $K$ such that,
for all processors $p_{i}$ and $p_{j}$
(including the case $i=j$),
perfectly reliable one-way communication from $p_{i}$ to $p_{j}$ in exactly $T$ communication rounds is achievable
whenever $p_{i}$ is non-faulty throughout the first $K$ of these rounds 
and $p_{j}$ is non-faulty throughout the last $K$ of these rounds.

Suppose furthermore that \\ 
(i) $K < \frac{n}{6m}$; \\
and (ii) there exists a value $R$ such that processor $p_{R}$ is non-faulty in all rounds from $(2R-2)T-K+1$ to $(2R-2)T+K$ inclusive
(the rounds $(2R-2)T-K+1$ to $(2R-2)T$ should be ignored if $R=1$). 

Then there exists a protocol for non-authenticated mobile Byzantine agreement such that for every $r \geq 2R$ \\
(a) all processors which are non-faulty during all rounds from $rT-K+1$ to $rT$ inclusive
will have a common value for $v_{s}^{\prime}$ at the end of round $rT$
and will not change it unless they become faulty; \\
and (b) if the source processor is non-faulty during all rounds from $1$ to $K$ inclusive
(i.e.~if $R=1$),
then this value of $v_{s}^{\prime}$ will always be the correct source value $v_{s}$.
\end{Theorem}
\textbf{Proof}
Note that at most $mK$ senders can be faulty at some stage during the first $K$ rounds of a communication procedure,
and that at most $mK$ receivers can be faulty at some stage during the last $K$ rounds of such a procedure.

The proof consequently follows from a careful examination of the protocol and proof for complete networks given in Section~\ref{latest}.
We use the exact structure of that protocol,
but with every communication round now replaced by $T$ rounds
and with the role of $m$ replaced by $mK$.
\phantom{qwerty}
\setlength{\unitlength}{.25cm}
\begin{picture}(1,1)
\put(0,0){\line(1,0){1}}
\put(0,0){\line(0,1){1}}
\put(1,1){\line(-1,0){1}}
\put(1,1){\line(0,-1){1}}
\end{picture} \\

If we are just interested in satisfying MBA$1^{\prime}$ and MBA2,
then we may omit some of the details from the statement of the theorem
and obtain the following simplified version:

\begin{Corollary} \label{maincor}
Let $m$ denote the number of mobile faulty processors in each communication round,
let $n$ denote the total number of processors,
and let $p_{1}, p_{2}, \ldots, p_{n}$ denote the processors themselves.

Suppose that there exist $T$ and $K$ such that,
for all processors $p_{i}$ and $p_{j}$
(including the case $i=j$),
perfectly reliable one-way communication from $p_{i}$ to $p_{j}$ in exactly $T$ communication rounds is achievable
whenever $p_{i}$ is non-faulty throughout the first $K$ of these rounds 
and $p_{j}$ is non-faulty throughout the last $K$ of these rounds.

Then if $K < \frac{n}{6m}$,
there exists a protocol that always satisfies conditions MBA$1^{\prime}$ and MBA2 for non-authenticated mobile Byzantine agreement.
\end{Corollary}
\textbf{Proof}
Recall that condition MBA$1^{\prime}$ is vacuous
unless there exists a processor $p_{R}$ that is always non-faulty,
and so we may assume that condition (ii) in Theorem~\ref{commtoprotocol} is satisfied.
Thus, we attain outcomes (a) and (b) from Theorem~\ref{commtoprotocol},
which are sufficient to guarantee the fulfillment of MBA$1^{\prime}$ and MBA2
at the end of round $2nT$.
\phantom{qwerty}
\setlength{\unitlength}{.25cm}
\begin{picture}(1,1)
\put(0,0){\line(1,0){1}}
\put(0,0){\line(0,1){1}}
\put(1,1){\line(-1,0){1}}
\put(1,1){\line(0,-1){1}}
\end{picture} \\

It now suffices to construct a communication system satisfying the conditions of Theorem~\ref{commtoprotocol}/Corollary~\ref{maincor},
and this will consequently be the focus of Section~\ref{apps}.

\section{Applications of the key theorem} \label{apps}

In the previous section,
we reduced the agreement problem to one of obtaining a system of reliable communication.
In this section,
we shall now construct suitable communications procedures,
consequently obtaining conditions for the existence of successful protocols
in terms of the minimum degree and vertex-connectivity of the network.
In Subsection~\ref{delta},
we derive a tight sufficient bound of $\delta > \frac{n}{2} + 2m - 1$ on the minimum degree
(see Theorem~\ref{mindeg});
and in Subsection~\ref{kappa},
we obtain a sufficient condition of $\kappa \geq 10m$ on the vertex-connectivity
(see Corollary~\ref{10mcor}).

\subsection{Results in terms of the minimum degree} \label{delta}

In this subsection,
we shall use transmissions through paths of length two to construct a method for reliable communication (Lemma~\ref{deltalemma}),
and consequently obtain a successful agreement protocol (Theorem~\ref{mindeg}).
This will all be phrased in terms of the minimum degree of the network,
but we shall note (in Corollary~\ref{deltatokappa})
that this also immediately implies results in terms of the vertex-connectivity.

We start with our result on reliable communication,
which satisfies the condition required in Theorem~\ref{commtoprotocol}/Corollary~\ref{maincor}
by taking $T=2$ and $K=1$.

\begin{Lemma} \label{deltalemma}
Let $\delta$ denote the minimum degree of the network,
let $m$ denote the number of mobile faulty processors in each communication round,
and let $n$ denote the total number of processors.
Then if $\delta > \frac{n}{2} + 2m-1$,
perfectly reliable one-way communication from any processor $u$ to any processor $v$ is achievable
in exactly $2$ communication rounds
as long as $u$ is non-faulty in the first of these rounds
and $v$ is non-faulty in the second.
\end{Lemma}
\textbf{Proof}
Note firstly that the case $u=v$ is trivial,
since the conditions then imply that $u$ is non-faulty in both rounds,
and so he can simply store any desired message to himself in his memory without fear of it being tampered with.
In the rest of the proof,
we shall deal with the case $u \neq v$.

Let $d(u)$ and $d(v)$, respectively,
denote the degrees of $u$ and $v$,
and observe that we have $d(u)+d(v) \geq n+4m-1$.
This inequality immediately implies that $u$ and $v$ must have at least $4m-1$ neighbours in common.
Furthermore,
if $u$ and $v$ are not themselves neighbours,
then
(since the size of the union of the neighbourhoods of $u$ and $v$
can then only be at most $n-2$)
the inequality implies that $u$ and $v$
must actually have at least $4m+1$ neighbours in common.

We may consequently always send a message from $u$ to $v$ through $4m+1$ different routes:
if $u$ and $v$ are not neighbours,
then we send the message along $4m+1$ pre-agreed paths of length $2$;
if $u$ and $v$ are neighbours,
then we send the message along $4m-1$ pre-agreed paths of length $2$,
and also directly from $u$ to $v$ in round $1$
(i.e.~the `path' $uvv$)
and directly from $u$ to $v$ again in round $2$
(i.e.~the `path' $uuv$).

If $u$ is non-faulty in round $1$ and $v$ is non-faulty in round $2$,
then since the intermediary nodes are all distinct,
it can then be observed that only at most $2m$ of the $4m+1$ transmissions may be disrupted
(at most $m$ in each round),
and so $v$ can simply use `majority vote' to obtain the correct message. 
\phantom{qwerty}
\setlength{\unitlength}{.25cm}
\begin{picture}(1,1)
\put(0,0){\line(1,0){1}}
\put(0,0){\line(0,1){1}}
\put(1,1){\line(-1,0){1}}
\put(1,1){\line(0,-1){1}}
\end{picture} \\

As an aside,
we note that the proof of Lemma~\ref{deltalemma} actually only requires that any pair of processors have at least $4m+1$ common neighbours
(or $4m-1$ common neighbours if the pair are themselves adjacent),
rather than that $\delta > \frac{n}{2} + 2m-1$.

Lemma~\ref{deltalemma} and Theorem~\ref{commtoprotocol}/Corollary~\ref{maincor} 
may now be combined to provide a successful agreement protocol.
For simplicity,
we present the result in an analogous form to Corollary~\ref{maincor},
although a more detailed result akin to Theorem~\ref{commtoprotocol} could also be obtained.

\begin{Theorem} \label{mindeg}
Let $\delta$ denote the minimum degree of the network,
let $m$ denote the number of mobile faulty processors in each communication round,
and let $n>6m$ denote the total number of processors.
Then if $\delta > \frac{n}{2}+2m-1$,
there exists a protocol satisfying conditions MBA$1^{\prime}$ and MBA2 for non-authenticated mobile Byzantine agreement.
Furthermore,
this bound is optimal,
in the sense that there exist networks with $\delta = \frac{n}{2}+2m-1$
for which no such protocol is possible.
\end{Theorem}
\textbf{Proof}
The existence of a successful protocol when $\delta > \frac{n}{2}+2m-1$
follows from Lemma~\ref{deltalemma} and Corollary~\ref{maincor}.

To see that this bound is optimal,
simply consider the network formed by taking two complete graphs each of order $\frac{n}{2}-2m$,
selecting the source processor from amongst these,
and then adding in $4m$ extra processors that are joined to everything.
It can be observed that the constructed network will have $\delta = \frac{n}{2}+2m-1$,
but also that the extra processors will form a cut-set of size $4m$
and hence (by Theorem~\ref{impossible4m}) that no successful protocol is possible.
\phantom{qwerty}
\setlength{\unitlength}{.25cm}
\begin{picture}(1,1)
\put(0,0){\line(1,0){1}}
\put(0,0){\line(0,1){1}}
\put(1,1){\line(-1,0){1}}
\put(1,1){\line(0,-1){1}}
\end{picture} \\

Since the minimum degree of the network is certainly always at least the vertex-connectivity,
we immediately obtain the following corollary to Theorem~\ref{mindeg}:

\begin{Corollary} \label{deltatokappa}
Let $m$ denote the number of mobile faulty processors in each communication round,
let $n>6m$ denote the total number of processors,
and let $\kappa > \frac{n}{2}+2m-1$ denote the vertex-connectivity of the network.
Then there exists a protocol satisfying conditions MBA$1^{\prime}$ and MBA2 for non-authenticated mobile Byzantine agreement.
\end{Corollary}

Note that Theorem~\ref{mindeg} also shows that there exist networks with vertex-connectivity only $4m+1$ for which successful protocols can be constructed
(consider the network formed by taking two complete graphs of order $\frac{n-4m-1}{2}$
and then adding in $4m+1$ extra vertices that are joined to everything),
and so the bound of Theorem~\ref{impossible4m} cannot be improved.

We shall produce further results in terms of the vertex-connectivity in the next subsection.

\subsection{Results in terms of the vertex-connectivity} \label{kappa}

At the end of the previous subsection,
we made some observations concerning successful agreement protocols in terms of $\kappa$,
the vertex-connectivity of the network.
In this subsection,
we shall now introduce a brand new method for reliable communication
(in Lemma~\ref{reliable}),
and then utilise Corollary~\ref{maincor} 
to consequently obtain further results
(Theorem~\ref{newcor} and Corollary~\ref{10mcor})
for the attainment of agreement protocols when $\kappa$ is sufficiently large.
We shall then also provide a result (Corollary~\ref{Amcor})
phrased for the case when the number of mobile faults is a specific proportion of the total number of processors.

We start with our new system for reliable communication,
which satisfies the condition required in Theorem~\ref{commtoprotocol}/Corollary~\ref{maincor}
by taking $K=T-1$:

\begin{Lemma} \label{reliable}
Let $m$ denote the number of mobile faulty processors in each communication round,
let $n$ denote the total number of processors,
let $\kappa > 4m$ denote the vertex-connectivity of the network,
and let $T = \lceil \frac{n-1-4m}{\kappa - 4m} \rceil >1$
(i.e.~the network is not complete).

Suppose that processor $u$ is non-faulty throughout communication rounds $1$ to $T-1$,
and that processor $v$ is non-faulty throughout communication rounds $2$ to $T$.
Then perfectly reliable one-way communication from $u$ to $v$ may be achieved in exactly $T$ rounds.
\end{Lemma}
\textbf{Proof}
Again,
we note that the case $u=v$ is trivial,
and so we may assume $u \neq v$.

If $T=2$,
then $n-1-4m \leq 2(\kappa - 4m)$,
which corresponds exactly to $\kappa \geq \frac{n}{2} + 2m - \frac{1}{2}$.
Thus, we certainly must have
$\delta \geq \frac{n}{2} + 2m - \frac{1}{2}$,
and so we are done,
by Lemma~\ref{deltalemma}.

Now let us consider the case $T \geq 3$.
If $u$ and $v$ are neighbours,
then they can simply exchange the message reliably in round $2$,
so we shall assume instead that all paths from $u$ to $v$ have length at least $2$.

By Menger's Theorem,
we know that there exists a set of $\kappa$ vertex-disjoint paths from $u$ to $v$
(we may assume that these are pre-agreed upon).
For all rounds from $1$ to $T-1$, inclusive,
let $u$ send the message along all $\kappa$ of these paths
(hence, if a given path has length $l$,
messages sent along that path will be received by $v$ from round $l$ onwards,
and some may not reach $v$ by time $T$).

Observe that $u$ sends the message $\kappa (T-1)$ times in total
($T-1$ times along each path),
and note that at most $n-2- \kappa$ of these transmissions will not have had time to reach $v$ by the end of round $T$
(this follows from observing that the points reached by such transmissions will all be distinct processors,
and will not be $u$ or $v$ or the first vertex in any of the $\kappa$ paths).
Hence, $v$ will receive at least $\kappa T-n+2$ transmissions in total.

When an intermediate node is faulty in a particular round,
this will affect both the message that it sends in that round and also the message that it stores
(and then sends in the following round).
Hence, since there will be $mT$ times that processors will be faulty,
and since we may ignore transmission faults in round $1$ and storage faults in round $T$,
the total number of messages that can be disrupted by the adversary is $2mT-m-m=2m(T-1)$.

We shall now observe that processor $v$ may then use `majority vote' to determine the correct message ---
it only remains to show that the number of disrupted messages is less than half the total number received,
i.e.~$2m(T-1) < \frac{\kappa T-n+2}{2}$.
But this is exactly equivalent to $T> \frac{n-2-4m}{\kappa -4m}$,
and so $T= \lceil \frac{n-1-4m}{\kappa - 4m} \rceil$ is certainly sufficient.
\phantom{qwerty}
\setlength{\unitlength}{.25cm}
\begin{picture}(1,1)
\put(0,0){\line(1,0){1}}
\put(0,0){\line(0,1){1}}
\put(1,1){\line(-1,0){1}}
\put(1,1){\line(0,-1){1}}
\end{picture} \\

The following result now follows immediately from Lemma~\ref{reliable} and Corollary~\ref{maincor}
(again, we also note that further details can be derived via Theorem~\ref{commtoprotocol} if desired):

\begin{Theorem} \label{newcor}
Let the network be non-complete,
let $m$ denote the number of mobile faulty processors in each communication round,
let $n$ denote the total number of processors,
and let $\kappa >4m$ denote the vertex-connectivity of the network.
Then if $\lceil \frac{n-1-4m}{\kappa - 4m} \rceil -1 < \frac{n}{6m}$,
there exists a protocol that always satisfies conditions MBA$1^{\prime}$ and MBA2 for non-authenticated mobile Byzantine agreement.
\end{Theorem}

We may also simplify Theorem~\ref{newcor} to obtain the following bound:

\begin{Corollary} \label{10mcor}
Let the network be non-complete,
let $m$ denote the number of mobile faulty processors in each communication round,
let $n$ denote the total number of processors,
and let $\kappa > 4m$ denote the vertex-connectivity of the network.
Then if $\kappa \geq 10m - 24 \frac{m^{2}}{n} - 6 \frac{m}{n}$,
there exists a protocol that always satisfies conditions MBA$1^{\prime}$ and MBA2 for non-authenticated mobile Byzantine agreement.
\end{Corollary}
\textbf{Proof}
In order to satisfy $\lceil \frac{n-1-4m}{\kappa - 4m} \rceil -1 < \frac{n}{6m}$,
it suffices if $\frac{n-1-4m}{\kappa - 4m} \leq \frac{n}{6m}$,
which is exactly equivalent to $\kappa \geq 10m - 24 \frac{m^{2}}{n} - 6 \frac{m}{n}$.
\phantom{qwerty}
\setlength{\unitlength}{.25cm}
\begin{picture}(1,1)
\put(0,0){\line(1,0){1}}
\put(0,0){\line(0,1){1}}
\put(1,1){\line(-1,0){1}}
\put(1,1){\line(0,-1){1}}
\end{picture} \\

Finally, we may also deduce our aforementioned result for the case when $\frac{n}{m} = A$:

\begin{Corollary} \label{Amcor}
Let $m$ denote the number of mobile faulty processors in each communication round,
let $n=Am$ denote the total number of processors
(for some $A>6$),
and let $\kappa$ denote the vertex-connectivity of the network.
Then if
\begin{displaymath}
\kappa \geq
\left\{ \begin{array}{ll}
(\frac{A}{2} + 2)m & \textrm{if $6<A\leq12$} \\
(10 - \frac{24}{A})m & \textrm {if $A \geq 12$}, \\
\end{array} \right. 
\end{displaymath} 
there exists a protocol that always satisfies conditions MBA$1^{\prime}$ and MBA2 for non-authenticated mobile Byzantine agreement.
\end{Corollary}
\textbf{Proof}
The $A \geq 12$ part of the result follows from Corollary~\ref{10mcor}
(note that the $\kappa > 4m$ condition is irrelevant here),
while the $A \leq 12$ part follows from Corollary~\ref{deltatokappa}.
\phantom{qwerty}
\setlength{\unitlength}{.25cm}
\begin{picture}(1,1)
\put(0,0){\line(1,0){1}}
\put(0,0){\line(0,1){1}}
\put(1,1){\line(-1,0){1}}
\put(1,1){\line(0,-1){1}}
\end{picture} \\

Note that this bound on $\kappa$ will be close to $5m$ when $A$ is close to $6$,
and will never exceed $10m$.

\section{Concluding remarks} \label{misc}

In this paper,
we have investigated the topic of agreement protocols in the presence of a mobile adversary,
looking at both the case when the communications network is complete
and when it is not complete.

For the former,
we obtained results (in Section~\ref{latest})
in terms of $n$,
the total number of processors,
and $m$,
the number of mobile faulty processors,
producing a successful protocol for the case when $n>6m$
and showing that $n>5m$ is certainly necessary.
For the latter,
we derived (in Sections~\ref{intro2} and~\ref{apps})
a tight bound of $\delta > \frac{n}{2} + 2m-1$ for a sufficient minimum degree,
and showed that vertex-connectivity $\kappa \geq 10m$ is also sufficient,
while $\kappa > 4m$ is certainly necessary.

It would be interesting to see whether the remaining gap between the necessary and sufficient conditions in the complete case can be closed.
Similarly,
it would be nice to obtain a tight result for the non-complete case in terms of the vertex-connectivity.
In particular,
do there exist any networks with vertex-connectivity greater than $4m$ for which agreement protocols are not possible?

\section*{Acknowledgments}
I would like to thank Stephen Wolthusen for introducing me to agreement protocols.

\end{document}